# Speed and accuracy in a visual motion discrimination task as performed by rats

**_Pamela Reinagel[1,*], Emily Mankin[2] and Adam Calhoun[2]_**

**_1. Neurobiology Section, Division of Biology, University of California San Diego_**

**_2. Department of Neurosciences, University of California San Diego_**

**_*to whom correspondence should be addressed: preinagel@ucsd.edu_**

## Abstract

**We find that rats, like primates and humans, perform better on the random dot motion task when they take more time to respond. We provide evidence that this improvement is due to stimulus integration. Rats increase their response latency modestly as a function of trial difficulty. Rats can modulate response latency more strongly on a trial by trial basis, apparently on the basis of reward-related parameters.**

## 1. Introduction

In the visual random dot motion task, subjects discriminate which of two directions the majority of dots are moving. The signal to noise ratio, and thus difficulty, is varied by the fraction of the dots moving coherently vs. randomly.

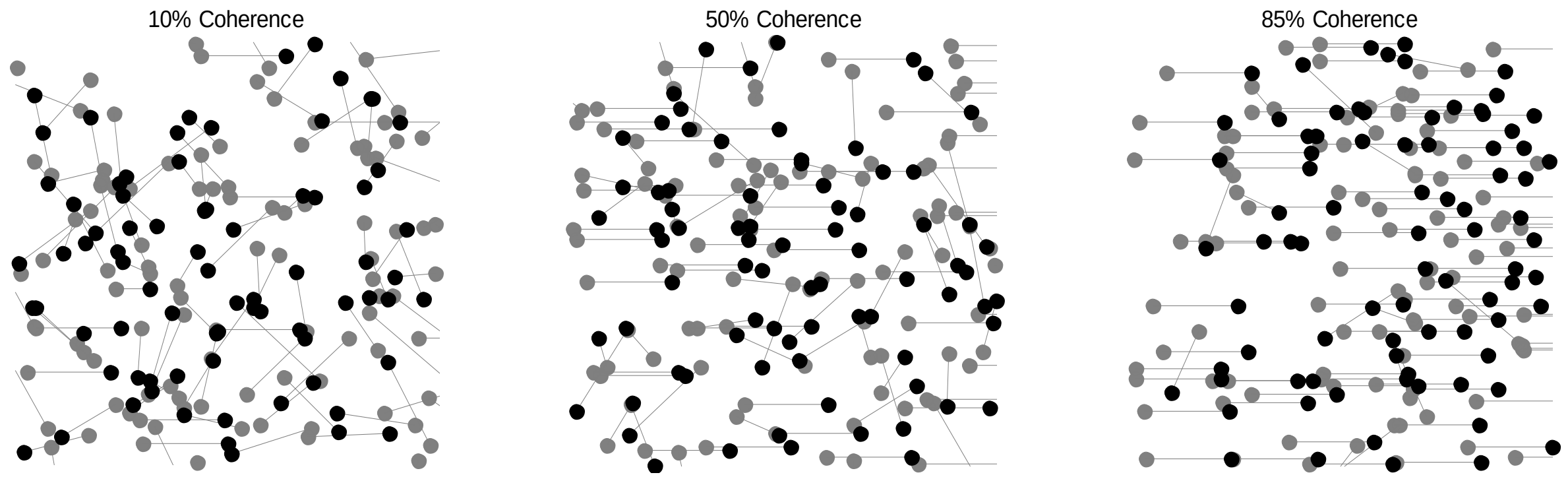

**Figure 1.1. Stimulus Paradigm.**

In primates, reaction time is longer and accuracy is lower in more difficult trials; for a given difficulty, increasing reaction time increases accuracy (Palmer et al 2005). It remains unclear whether the same is true of rodents, at least in other perceptual tasks (Uchida & Mainen 2003; Rinberg et al 2006). In order to compare perceptual decision making of primates and rats more directly we trained rats to perform the identical visual motion task studied in primates.

This preliminary report is identical in content to a poster presented at the 2009 Society for Neuroscience meeting. The content differs from the published meeting abstract (Reinagel et al, 2009) in that the result of Figure 3.3 contradicts and supersedes the claim to the contrary in the abstract, and the results of Figures 3.4, 3.5 and 3.6 were not mentioned in the abstract. This document differs from the poster (Appendix 1) in that a brief narrative explanation of each figure has been added.

## 2. Methods

Freely behaving, water-restricted rats performed a 2-alternative forced-choice motion discrimination task for water reward. Each time the subject requested a trial, 100 white dots were drawn against a black background on a CRT monitor. Dot size and motion speed were chosen to be approximately optimal for rats. A percentage of the dots moved in a coherent direction (either left or right); the remaining dots moved randomly. The coherence was chosen independently each trial. The motion stimulus continued until the subject responded, with no time limit. The task was to lick the response/reward port located on the side towards which the coherent dots moved. Correct responses earned small water rewards; incorrect responses were penalized by a 2 second time-out before a new trial could be requested. We define the “response latency” as the time between the trial-initiating request and the trial-terminating response.

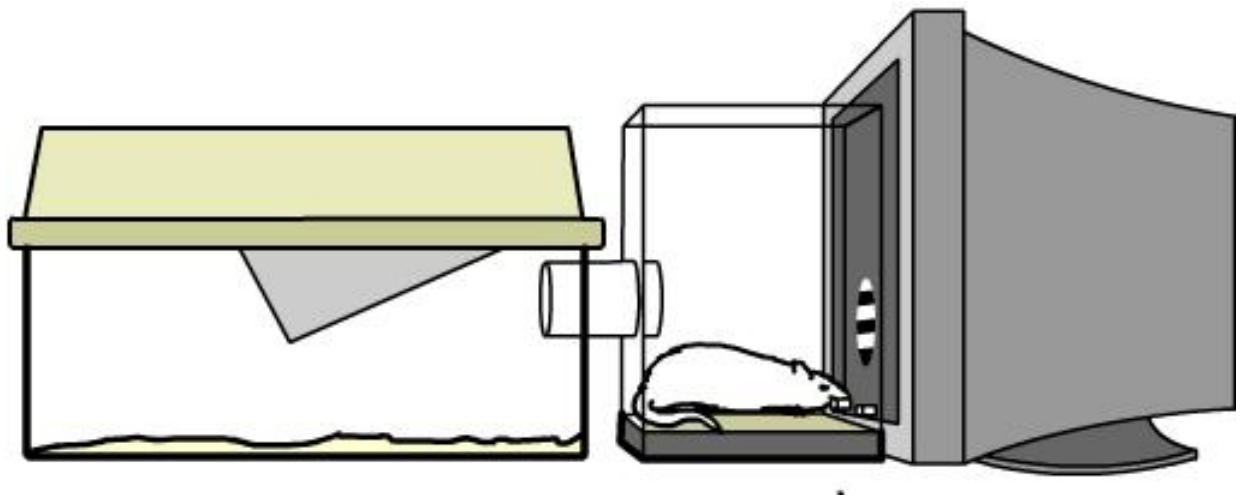

**Figure 2.1. Training Apparatus**

Rats typically perform hundreds of trials per 2 hour session, one session per day, 7 days a week. Naïve p30 rats were initially trained using 85% coherence, for which they reached 85% correct criterion within a few days. Coherence was then varied from trial to trial to modulate task difficulty.  We recorded reaction time and trial outcome daily in an automated training and testing apparatus attached to the home cage.

We report here the results from 2 male Long-Evans rats. Analysis was restricted to training blocks of >=1000 consecutive trials during which the stimulus distribution was fixed and performance was stationary.

*Additional Training Details*

We used stochastic correction trials in order to discourage bias or perseveration. Therefore trials immediately following incorrect responses were not included in analysis, whether or not a correction trial was given.

Reward amplitude in each trial was ramped, increasing with the number in a row correct. This was designed to make random guessing a lean task strategy. A consequence is that after a long winning streak, the stakes of a trial are high: correct answers get large rewards and perpetuate the high-reward regime, while wrong answers reset the ramp to the lowest reward value in the next trial. Time out duration was not ramped, but remained fixed at 2 seconds.

In some blocks, the duration of the motion stimulus was pre-determined each trial by drawing a value uniformly between 25ms and 250ms. The stimulus began when subject requested a trial, and continued for the predetermined time regardless of behavioral response. There was no time limit to respond. These data are analyzed in Figures 3.4 and 3.5.

# 3. Results

## Accuracy increases with motion strength

As expected, accuracy in discriminating the direction of coherent motion was dependent on the coherence level (Figure 3.1). Near 0% coherence, performance was at chance. At 100% coherence this subject reached just over 80% correct. The just-detectable coherence was in the range of 10-20% coherence, which is much higher than typically reported for primates. However we note that the size of the dots relative to the acuity of the rat is probably much smaller than would be the case in most human and non-human primate studies, so this comparison may be misleading. Qualitatively similar results were found for a second subject (not shown).

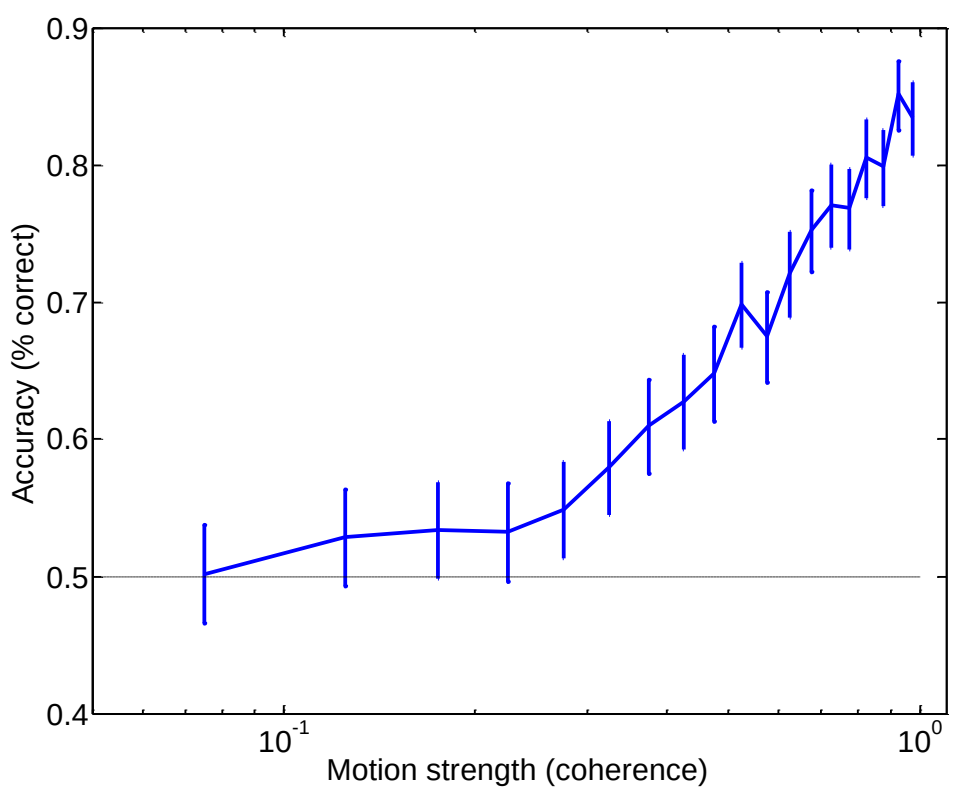

**Figure 3.1. Accuracy of motion discrimination for coherence ranging from 0 to 100%.** Performance is averaged over all response latencies. Data shown from one subject (r196) for 16,067 trials. Error bars indicate 95% confidence bounds for binomial distribution. Similar results obtained in several training blocks with different task parameters, and for one additional subject.

## Accuracy increases with response latency

In this task there was no response deadline, and the stimulus continued to be refreshed until the subject responded. Response latency (and thus stimulus duration) was typically less than 2 seconds for both subjects. We determined the accuracy of discrimination (% correct responses) as a function of both coherence and response latency. We found that for any given coherence, accuracy increases with response latency (Figure 3.2A). Note that when coherence is high (>=85%), this subject reached 100% correct performance for response latencies over 500ms. Thus the lapse for 85% coherence seen in Figure 3.1 can be explained by failures in trials with very short response latency. At weaker the coherences, performance increased to an asymptotic value less than 100%; the latency at which this asymptote occurred, as well as the asymptotic performance level, were monotonically related to coherence (Figure 3.2A).

Performance was at chance when response latency was <200ms regardless of coherence (Figure 3.2B), indicating that these very short responses (which were rare) are not stimulus-dependent. But accuracy improved with stimulus coherence for reaction times of 200-300ms (green curve, Figure 3.2B), implying that rats were relying on the visual stimulus at least in part to make their behavioral choice even for very rapid responses. Reaction times of 500-800ms were sufficient to achieve asymptotic performance at all coherence levels (magenta vs. black curve, Figure 3.2B).

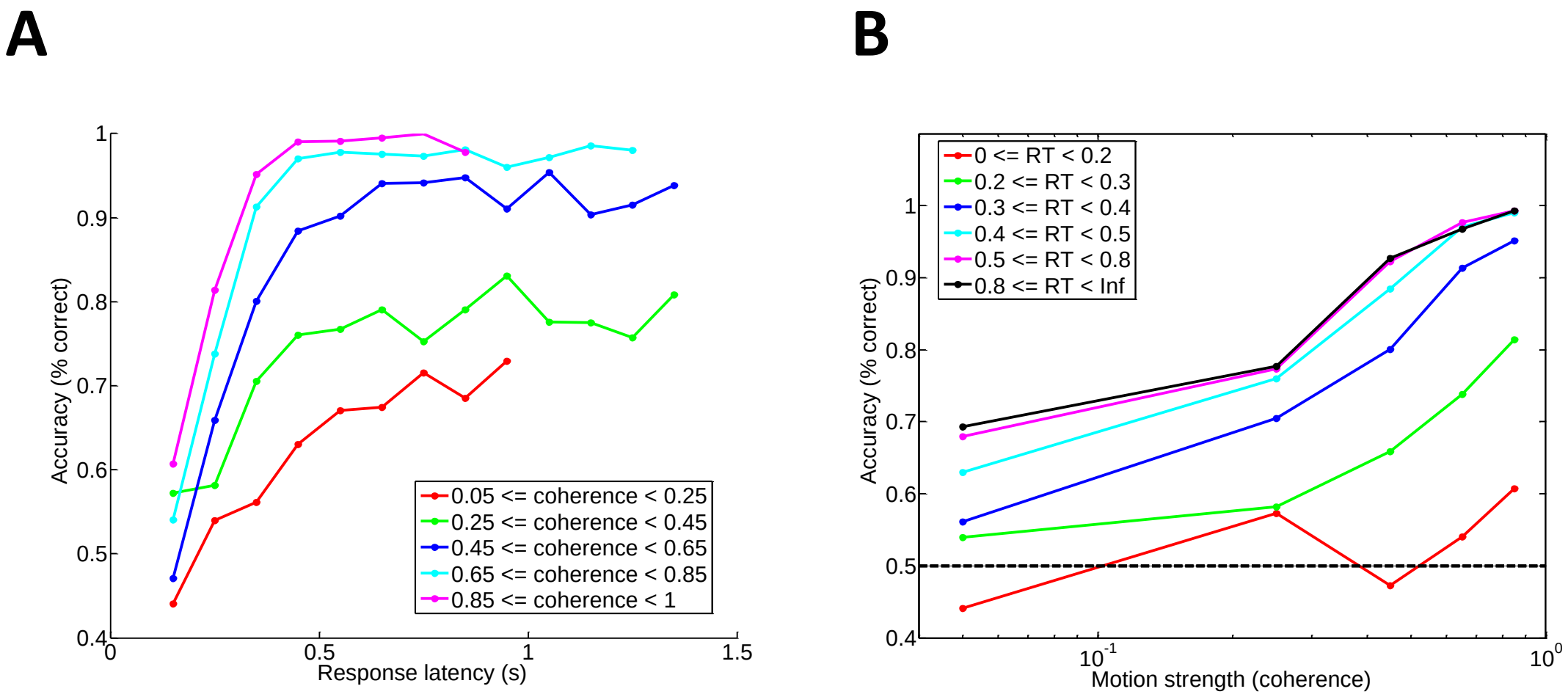

**Figure 3.2. Accuracy higher when response later. (A)** Accuracy of motion discrimination as a function of response latency, plotted separately for different coherence ranges. Data shown from one subject (r196) for 29,261 trials during which coherence ranged from 20% to 90%. Error bars indicate 95% confidence bounds for binomial distribution. Similar results observed for one additional subject. **(B)** Same data re-plotted as in Fig 2.1 for several latency ranges.

## Rats wait longer when motion is weaker

We found a modest but consistent shift in the reaction time distribution as a function of stimulus coherence. As coherence decreased, reaction time increased, consistent with a strategy of integration of evidence with time to overcome noise. For the example shown (Figure 3.3) the subject waited 19% longer to respond in low signal-to-noise ratio trials (20-25% coherence) compared with high SNR trials (85-90% coherence).

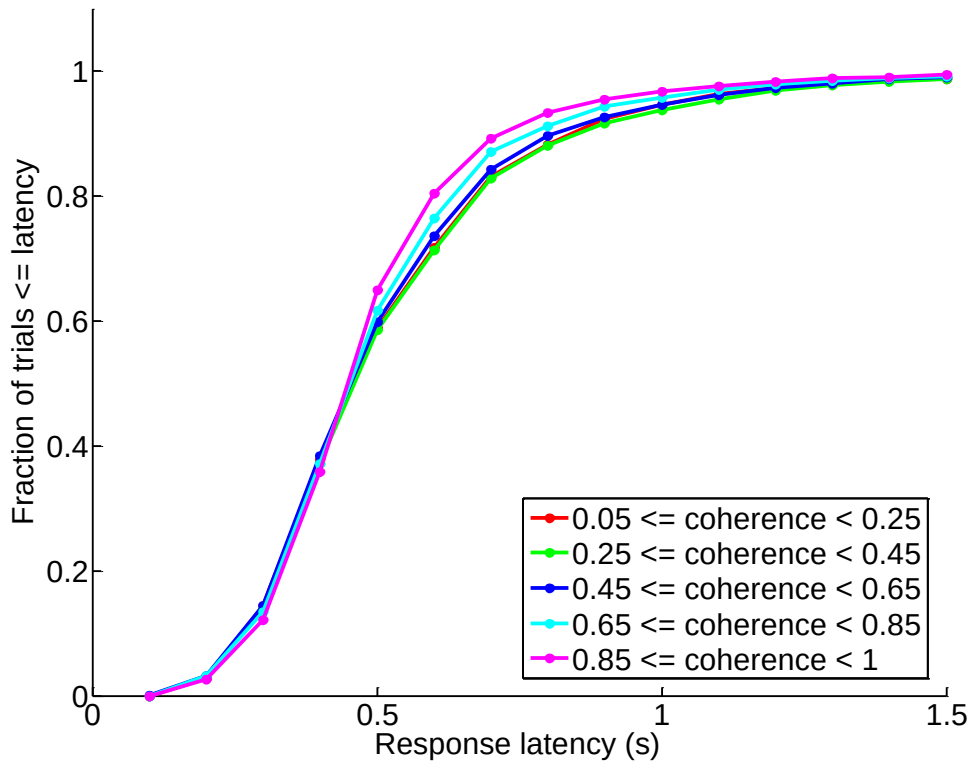

**Figure 3.3. Rats wait longer when motion is weaker.** Cumulative distribution of response latency for same trials shown in Figure 3.2, separated by coherence. There is a small but consistent shift to the right as motion coherence decreases. For this rat in this training block, the 90th percentile response latency time increased from 722 ms (for 85-90% coherence) to 860 ms (for 20-25% coherence). Similar results found in other training blocks and for one additional subject.

## Accuracy depends on stimulus duration.

The improvement in accuracy with response latency (Figure 3.2A) suggests that stimulus information is integrated over time. However those data could not exclude alternative interpretations. For example, both accuracy and response latency might both correlate with some hidden variable, such as motivation or attention. To disambiguate these, we performed an additional experiment in which the stimulus duration was under experimental control (see Methods). The task was still self-paced: the stimulus appeared at the time of voluntary request, persisted for a predetermined time, followed by a uniform black screen until the rat made a trial-terminating response. We found that for 85% coherence stimuli, discrimination accuracy increases with stimulus duration over the range 25-200ms (Figure 3.4). This result is consistent with the interpretation that stimulus integration per se underlies the performance improvements with response latency described above.

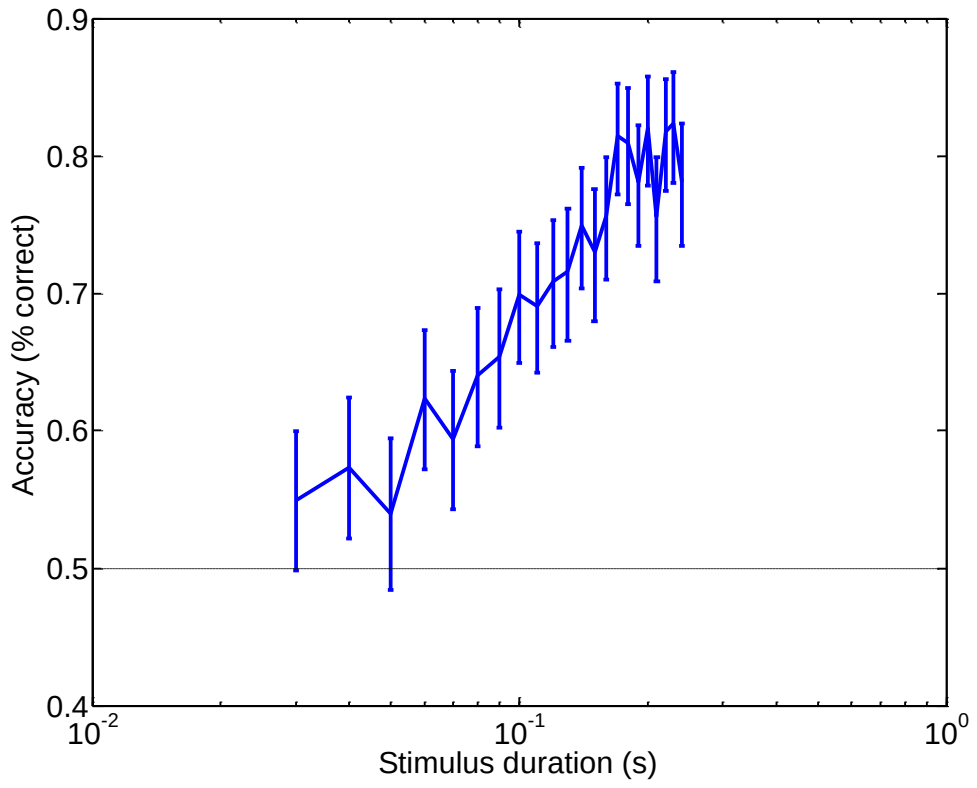

**Figure 3.4. Accuracy depends on stimulus duration.** Accuracy of motion discrimination for 85% coherence, as a function of stimulus duration ranging from 25ms to 250ms. Duration was randomly chosen each trial, independent of behavioral response. Performance is averaged over all response latencies. Data shown from one subject (r196) for 8295 trials. Error bars indicate 95% confidence bounds for binomial distribution. Similar results observed for one additional subject.

## Accuracy falls with time after offset

The dependence of accuracy on stimulus duration confirms a role for accumulation of sensory evidence in this task, but it does not exclude an additional role for cognitive processing time. Perhaps even after the stimulus extinguished, rats perform better when they take longer to process this information. Alternatively, the accumulated evidence may simply be degraded with neural noise over time after stimulus offset. In the experiment described in Figure 3.4, the maximum stimulus duration (250ms) was short compared to the minimum response latency, so stimuli were rarely terminated by responses. Consistent with the second interpretation, we find that for any given stimulus duration, accuracy generally falls off with response latency, decaying to chance within a few hundred milliseconds (Figure 3.5). There is some hint that briefer stimuli (for which less evidence accumulated) decay at a steeper rate, but we did not have sufficient data to establish this clearly. We note that the apparent improvement in accuracy after stimulus offset for the longest stimulus durations (magenta curve in Figure 3.5) was not consistent across subjects.

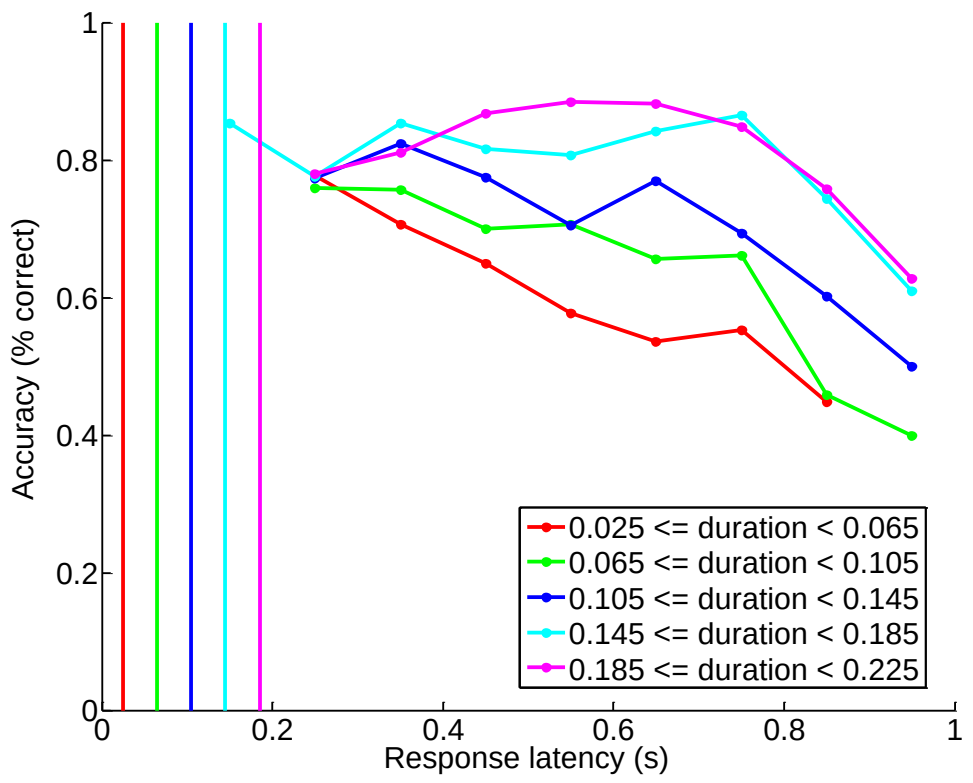

**Figure 3.5. Accuracy falls with time after offset.** Accuracy as a function of response latency, in the case when stimulus duration is limited (same data as in Figure 4). Each curve represents a range of stimulus durations; the latest corresponding stimulus offset time is indicated by the vertical line of same color. Similar results observed in this and other tasks for this and 1 additional subject.

## Rats can modulate response latency on a trial by trial basis

Above we showed that when rats control stimulus duration, they can voluntarily modulate their response latency on a trial by trial basis (Figure 3.2.). Rats choose to wait longer before responding when stimuli are more ambiguous, despite incurring a delay in the timing of the possible reward as well as a delay in the soonest possible next trial. This effect was rather weak however: there was at most 200ms difference in response latency between the weakest and strongest coherence stimuli. Given that rats seem to be able to integrate only 200ms of stimulus evidence (Figure 3.4), and the probability of reward is 50% at chance, however it is not clear that waiting longer would have been beneficial from the point of view of reward harvesting.

Therefore it is possible that rats' capacity to modulate response latency is greater than revealed by the effect of stimulus coherence. To test this we took advantage of a detail of our training protocol: in the interest of discouraging guessing, reward magnitude increased deterministically according to the number of correct responses in a row. Thus a strategy of stimulus-independent guessing will yield water reward in 50% of trials, but these rewards will typically be small. Large rewards occur after a string of successful trials, which are relatively rare if the rat is performing at chance.

Rats in principle could remember their recent success history, and thereby know in advance whether the available reward in the next trial will be large or small. Moreover, one incorrect response resets the reward in the subsequent trial to the smallest reward, beginning a new ramp. Therefore trials after a winning streak have higher stakes than trials after errors. The subjects were trained with ramped rewards over their entire lifespan, and therefore had ample opportunity to discover these rules.

If rats fail to predict the reward magnitude of the current trial, response latency should not depend on the size of the expected reward. If rats do make accurate predictions of expected reward magnitude, but fail to take into account the stakes of the trial, one might expect rats to respond faster when they know the reward is large. But we find that rats wait longer to respond on trials with higher expected rewards, (Figure 3.6), consistent with modulating their behavior to optimize reward over time. This effect can be quite large relative to the median response latency or the duration of demonstrable stimulus integration.

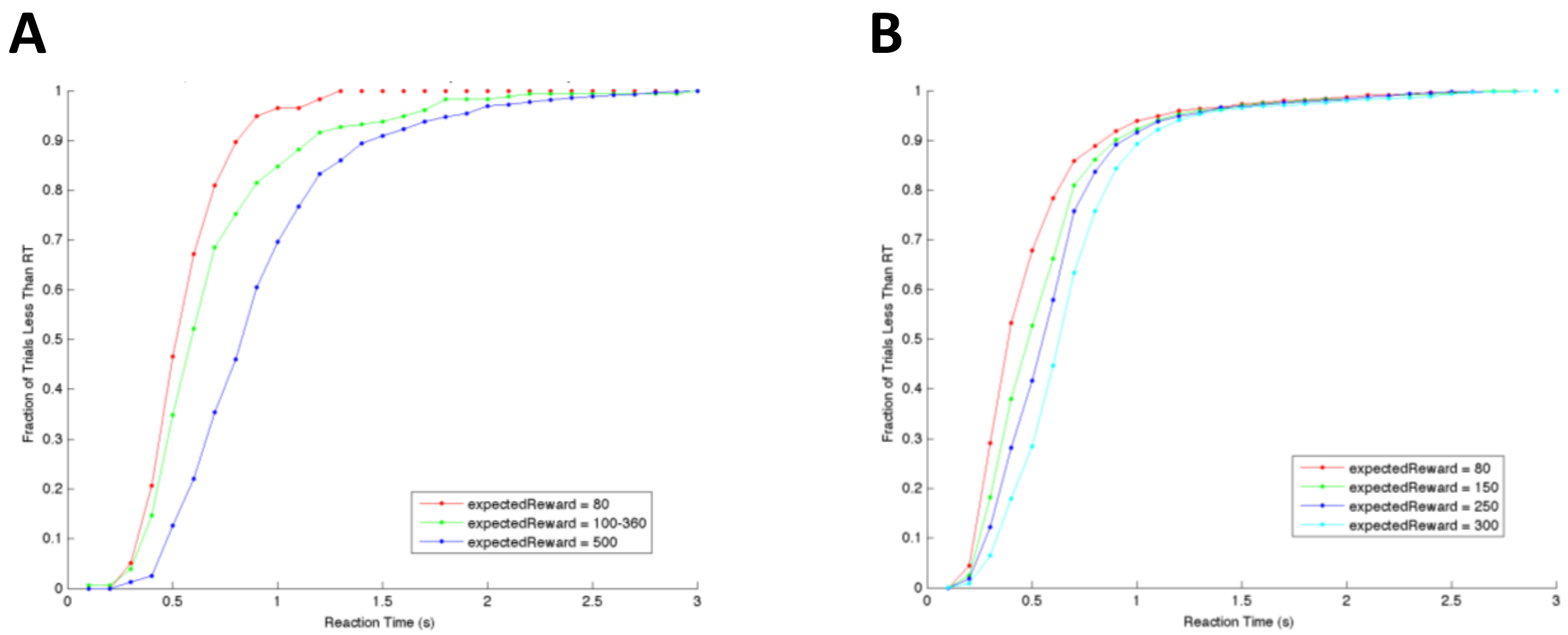

**Figure 3.6. Rats wait longer when stakes are higher. (A)** Cumulative distribution of response latencies when risk/reward is low (red), medium (green) and high (blue) due to reward ramping (see Methods details). Data shown for one rat, 829 trials from a training block in which the reward ramp was steep and stimulus parameters were constant (unlimited duration, 85% coherence). There is a substantial shift to the right (waiting longer) as the stakes increase. **(B)** The same result in another training block. The same results were also found for one additional subject.

## Discussion

Rats can perform the random dot coherent motion task that is widely used in primate and human studies of decision making. In our paradigm, rats' performance ranged from chance (50% correct) at low coherence to >85% correct at high coherence (Figure 3.1). For a given coherence, rats' accuracy increased with reaction time over the range 200-800ms. In long latency trials, performance is near 100% correct for high coherence stimuli (Figure 3.2). These results are consistent with the hypothesis that rats integrate sensory evidence over time in a similar manner to human and non-human primate subjects.

Rats' response latency increased with trial difficulty (Figure 3.3), waiting longer to respond when coherence was low. This result is consistent with the hypothesis that rats are able to modulate their response latency in order to trade off speed against accuracy.

Two lines of evidence suggest that the benefit of longer response latency is probably stimulus integration as opposed to other cognitive factors. First, accuracy increased with stimulus duration, up to ~200ms for 85% coherent stimuli (Figure 3.4). Second, accuracy decreased with response latency after stimulus offset (Figure 3.5), consistent with corruption of the accumulated sensory evidence with neural noise.

It remains to be determined whether the magnitude of the shift in response latency with coherence is optimal in the sense of maximizing reward harvesting over time. To fully model this we will need additional experiments. For example we will need to measure the dependence of accuracy on stimulus duration (Figure 3.4) for all coherence levels, and measure the temporal discounting time constant for each rat.

Finally we provide an existence proof that rats can modulate response latency strongly on a trial by trial basis, apparently in response to risk/reward expectations (Figure 3.6). Specifically, rats wait longer to decide when the stakes are high, despite the fact that the expected reward is large. Interpretation is limited by the fact that several reward parameters were linked in our experiment. The high stakes trials had larger available rewards on correct response, a higher cost for incorrect responses (greater reduction in reward magnitude for subsequent trials). Moreover, these trials occurred after winning streaks, and therefore the recent history of reward and stimuli are statistically different from the low-stakes trials. To untangle these factors we will need additional experiments in which the stakes of a trial are determined independently in each trial and subjects are cued on a trial by trial basis.

These results were obtained with only two subjects; additional subjects will be needed to fully validate and quantify these findings.

## Acknowledgements

This research was supported by grants from the Kavli institute of Mind and Brain at UCSD and a Scholar Award from the James S McDonnell Foundation. The authors are grateful to Erik Flister and Philip Meier for sharing the methods and technology used for automated training and testing (Meier, Flister and Reinagel, 2011) in advance of publication. Sarah Meder, Danielle Dickson, and Alee Wise provided animal care and oversight of training.

## Appendix 1. Replica of poster presented at the Society for Neuroscience Meeting, 2009.

# 281.12 EE118 Speed and accuracy in a visual motion discrimination task as performed by rats

*Pamela Reinagel[1], Emily Mankin[2] and Adam Calhoun[2]*

*1. Neurobiology Section, Division of Biology, and 2. Department of Neurosciences; University of California San Diego.*

UCSD

## Abstract

We find that rats, like primates and humans, perform better on the random dot motion task when they take more time to respond. We provide evidence that this improvement is due to stimulus integration. Rats increase their response latency modestly as a function of trial difficulty. Rats can modulate response latency more strongly on a trial by trial basis, apparently on the basis of reward-related parameters.

## Introduction

In the visual random dot motion task, subjects discriminate which of two directions the majority of dots are moving. The signal to noise ratio, and thus difficulty, is varied by the fraction of the dots moving coherently vs. randomly.

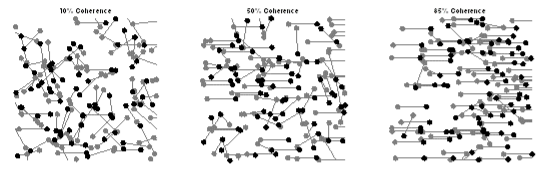

In primates, reaction time is longer and accuracy is lower in more difficult trials; for a given difficulty, increasing reaction time increases accuracy (Palmer et al 2005). It remains unclear whether the same is true of rodents, at least in other perceptual tasks (Uchida & Mainen 2003; Rinberg et al 2006). In order to compare perceptual decision making of primates and rats more directly we trained rats to perform the identical visual motion task studied in primates.

## Methods

Freely behaving, water-restricted rats performed a 2-alternative forced-choice motion discrimination task for water reward. Each time the subject requested a trial, 100 white dots were drawn against a black background on a CRT monitor. Dot size and motion speed were chosen to be approximately optimal for rats. A percentage of the dots moved in a coherent direction (either left or right); the remaining dots moved randomly. The coherence was chosen independently each trial. The motion stimulus continued until the subject responded, with no time limit. The task was to lick the response/reward port located on the side towards which the coherent dots moved. Correct responses earned small water rewards; incorrect responses were penalized by a 2 second time-out before a new trial could be requested. We define the "response latency" as the time between the trial-initiating request and the trial-terminating response.

Rats typically perform hundreds of trials per 2 hour session, one session per day, 7 days a week. Naïve p30 rats were initially trained using 85% coherence, for which they reached 85% correct criterion within a few days. Coherence was then varied from trial to trial to modulate task difficulty. We recorded reaction time and trial outcome daily in an automated training and testing apparatus attached to the home cage.

We report here the results from 2 male Long-Evans rats. Analysis was restricted to training blocks of >=1000 consecutive trials during which the stimulus distribution was fixed and performance was stationary.

**Additional Training Details**

We used stochastic correction trials in order to discourage bias or perseveration. Therefore trials immediately following incorrect responses were not included in analysis, whether or not a correction trial was given.

Reward amplitude in each trial was ramped, increasing with the number in a row correct. This was designed to make random guessing a lean task strategy. A consequence is that after a long winning streak, the stakes of a trial are high: correct answers get large rewards and perpetuate the high-reward regime, while wrong answers reset the ramp to the lowest reward value in the next trial. Time out duration was not ramped, but remained fixed at 2 seconds.

In some blocks, the duration of the motion stimulus was pre-determined each trial by drawing a value uniformly between 25ms and 250ms. The stimulus began when subject requested a trial, and continued for the predetermined time regardless of behavioral response. There was no time limit to respond. These data are analyzed in Figures 4 and 5.

## Results

### 1. Accuracy increases with motion strength

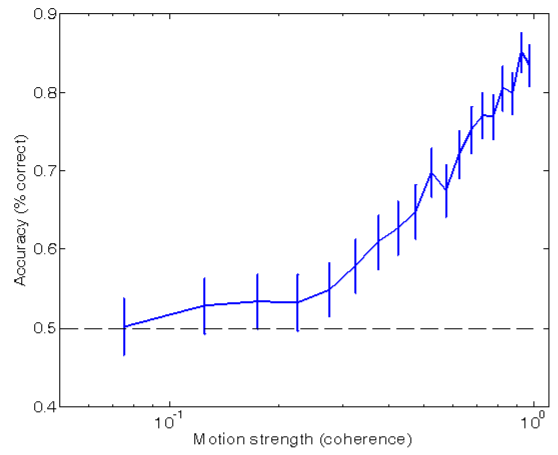

Accuracy of motion discrimination for coherence ranging from 0 to 100%. Performance is averaged over all response latencies. Data shown from one subject (r196) for 16,067 trials. Error bars indicate 95% confidence bounds for binomial distribution. Similar results obtained in several training blocks with different task parameters, and for one additional subject.

### 2. Accuracy higher when response later

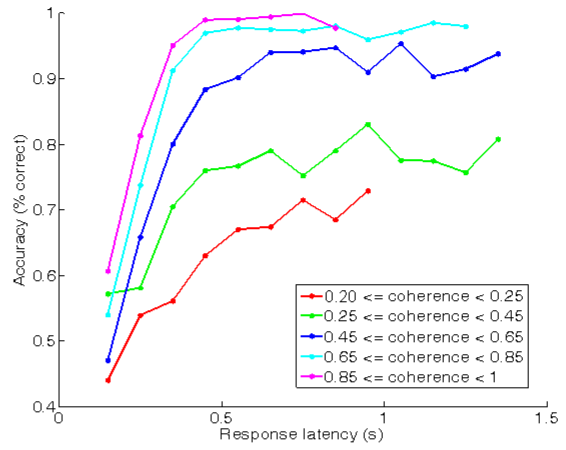

Accuracy of motion discrimination as a function of response latency, plotted separately for different coherence ranges. Data shown from one subject (r196) for 29,261 trials during which coherence ranged from 20% to 90%. Error bars indicate 95% confidence bounds for binomial distribution. Similar results observed for one additional subject. See also 2a below – same data re-plotted as in Fig 1 for several latency ranges.

### 3. Rats wait longer when motion is weaker

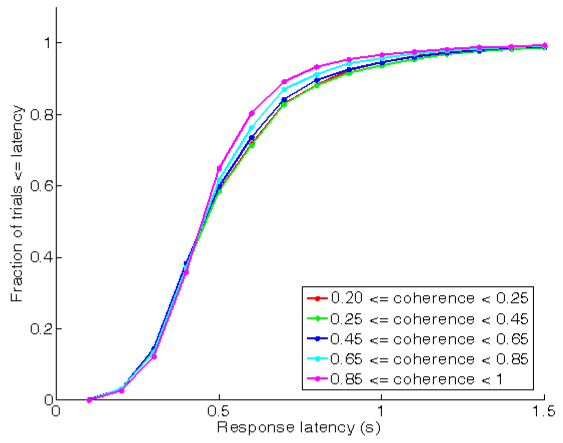

Cumulative distribution of response latency for same trials shown in Figure 2, separated by coherence. There is a small but consistent shift to the right as motion coherence decreases. For this rat in this training block, the 90th percentile response latency time increased from 722 ms (for 85-90% coherence) to 860 ms (for 20-25% coherence). Similar results found in other training blocks and for one additional subject.

### 4. Accuracy depends on stimulus duration

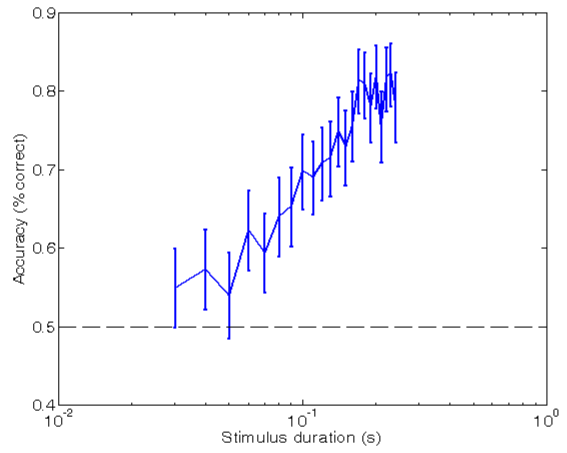

Accuracy of motion discrimination for 85% coherence, as a function of stimulus duration ranging from 25ms to 250ms. Duration was randomly chosen each trial, independent of behavioral response. Performance is averaged over all response latencies. Data shown from one subject (r196) for 8295 trials. Error bars indicate 95% confidence bounds for binomial distribution. Similar results observed for one additional subject.

### 5. Accuracy falls with time after offset

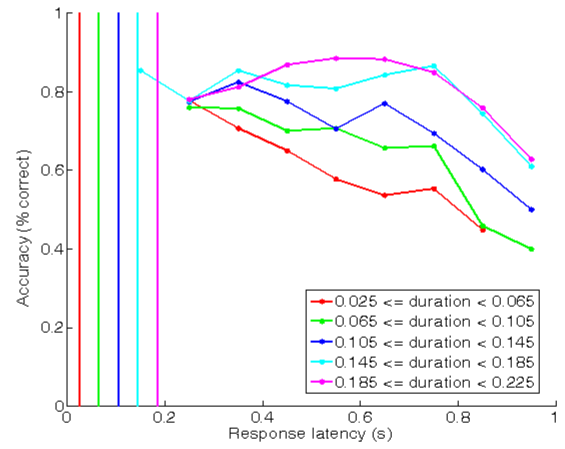

Accuracy as a function of response latency, in the case when stimulus duration is limited (same data as in Figure 4). Each curve represents a range of stimulus durations; the latest corresponding stimulus offset time is indicated by the vertical line of same color. Similar results observed in this and other tasks for this and 1 additional subject.

### 6. Rats wait longer when stakes are higher?

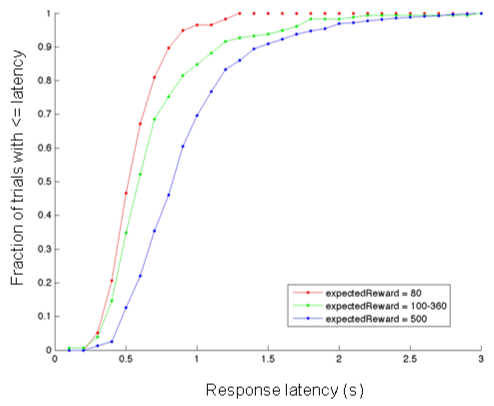

Cumulative distribution of response latencies when risk/reward is low (red), medium (green) and high (blue) due to reward ramping (see Methods details). Data shown for one rat, 829 trials from a training block in which the reward ramp was steep and stimulus parameters were constant (unlimited duration, 85% coherence). There is a substantial shift to the right (waiting longer) as the stakes increase. The same result was found in other training blocks (see also 6a below), and for one additional subject.

### Training Apparatus

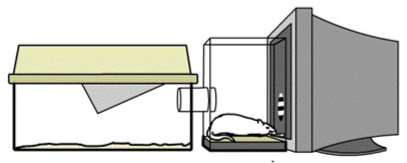

### Random walk model

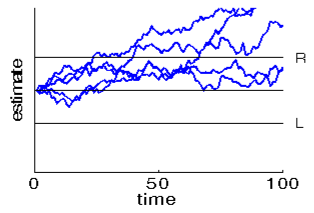

### 2a. Accuracy vs latency

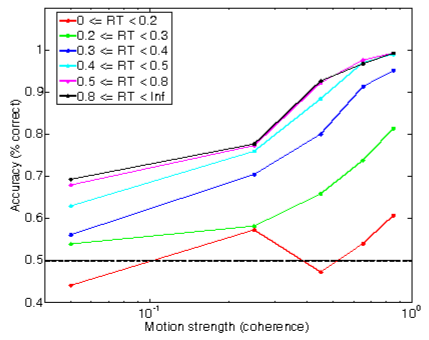

### 6a. Latency vs stakes

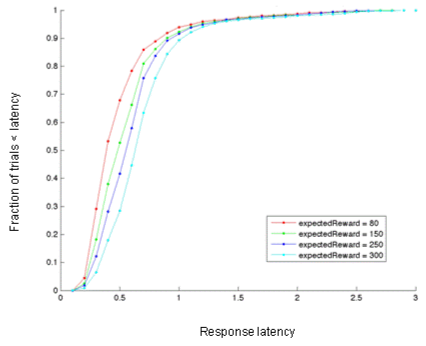

## Conclusions

(1) Rats' performance ranged from chance (50% correct) at low coherence to >85% correct at high coherence.
(2) For a given coherence, rats' accuracy increased with reaction time. In long latency trials performance is near 100% correct for high coherence stimuli.
(3) Response latency increased with trial difficulty eg., ~140ms difference in 90th percentile response latency between difficult and easy stimuli in the case shown.
(4) The benefit of longer response latency is probably stimulus integration, not cognitive processing
   a) performance increased with stimulus duration up to ~200ms
   b) performance decreased with latency after stimulus offset
(5) Rats can modulate response latency strongly on a trial by trial basis, apparently in response to risk/reward expectations.

Additional experiments are needed to increase N and to determine which of several linked reward parameter(s) affect response latency.

**Supported by**
Kavli institute of Mind and Brain at UCSD
James S McDonnell Foundation